\documentclass[doublecol]{epl2} 

\usepackage{graphicx}	
\usepackage{epstopdf}
\usepackage{dcolumn}	
\usepackage{bm}		
\usepackage{color}

\usepackage{amsmath}
\usepackage{amssymb}
\usepackage{amsfonts}
\usepackage{latexsym}

\usepackage{ulem}
\usepackage{color}

\usepackage[colorlinks,bookmarks=false,citecolor=blue,linkcolor=red,urlcolor=blue]{hyperref}

\newcommand{\bea}{\begin{eqnarray}}
\newcommand{\eea}{\end{eqnarray}}

\newcommand{\be}{\begin{equation}}
\newcommand{\ee}{\end{equation}}

\newcommand{\nn}{\nonumber \\}

\newcommand{\up}{\uparrow}
\newcommand{\down}{\downarrow}
\newcommand{\Leff}{L_{\mathrm{eff}}}
\newcommand{\TK}{T_{\mathrm{K}}}

\newcommand{\im}{{\mathrm{i}}}

\newcommand{\et}{{\it et al. }}

\title{Integrable impurities as boundary conditions}

\author{
     M. Moliner\inst{1,2} \thanks{E-mail: \email{Marion.Moliner@kit.edu}} 
\and P. Schmitteckert\inst{1,2} 
}
\shortauthor{M. Moliner and P. Schmitteckert}

\institute{                
  \inst{1} Institute of Nanotechnology, Karlsruhe Institute of Technology, 76344 Eggenstein-Leopoldshafen, Germany.\\ 
  \inst{2} Center of Functional Nanostructures, Karlsruhe Institute of Technology, 76131 Karlsruhe, Germany.
}


\abstract{
A few exactly solvable interacting quantum many-body problems with impurities were previously reported to exhibit unusual features such as non-localization and absence of backscattering. 
In this work we consider the use of these integrable impurities as boundary conditions in the framework of linear transport problems.
We first show that such impurities enhance the density of states at the Fermi surface, thus increasing the effective system size. 
The study of the real time-dynamics of a wave packet sent through a series of them inserted in both non-interacting and interacting leads 
then indicates that these impurities are transparent and do not add artefacts to the measurement of transport properties. 
We finally apply these new boundary conditions to study the conductance of an interacting scatterer using the embedding method. 
}

\begin{document}

\maketitle
\section{Introduction} 
The main difficulty in simulating strongly correlated quantum systems consists in the fact that strong correlations correspond to low energy, {\it i.e.}\ small temperature and large system size which leads to exponentially large Hilbert spaces. 
In his seminal work on the solution of the Kondo problem via the Numerical Renormalization Group technique (NRG)
Wilson~\cite{Wilson_1975} solved this problem by describing the lead or the bath coupled to an impurity by an effective
model, where he first introduced a logarithmic discretization in energy space, which he then transformed
into a nearest neighbour hopping chain via a Lanczos tridiagonalization. In this effective lead the hopping elements
are decaying exponentially with distance from the impurity leading to an exponential increase of the density of states
at zero energy. This approach has by now been established as the reference/benchmark method~\cite{Bulla_2008} to solve
Kondo like problems.
Motivated by this success the NRG was recently extended to describe non-equilibrium properties and even steady-state physics
of the Single Impurity Anderson model (SIAM)~\cite{Anders_2005}.
However, in the context of transport calculations the Wilson chain possesses one major disadvantage. Each change
in the hopping elements introduces a scatterer which leads to backscattering, compare~\cite{Schmitteckert_2010}. 
Thus, one has to be very careful, not to measure the transport properties of these "auxiliary" scatterers, especially if one is interested in on-resonance phenomena.
Here we apply integrable scatterers instead of Wilson's reduced hopping elements which share the property of
an exponentially enhanced density of states at zero energy, while not introducing any backscattering due to their integrable nature.

\begin{widetext}
\bea
 \mathcal{H}_j^{\Delta} = &-& V\sinh \nu_{j} \Big( 2\hat{n}_{j} \hat{n}_{j+1} - 2\hat{n}_{j-1} \hat{n}_{j+1} + 2\hat{n}_{j-1} \hat{n}_{j} - 2\hat{n_j} - (-1)^{\hat{n}_j} \big(c_{j-1}^{\dagger}c_{j+1}^{} + \mathrm{h.c.}  \big) \Big) 
		    \nn
		    &+&	2 \frac{(V^2/4 + \cosh \nu_j)(\cosh \nu_j - 1)}{\sinh \nu_j}\big( c_{j}^{\dagger}c_{j-1}^{} + c_{j}^{\dagger}c^{}_{j+1} + \mathrm{h.c.} \big)			
			 \label{eq:Ham_one_II} \\
		    &+& \im \sqrt{1-\frac{V^2}{4}} \Big(2\cosh \nu_j \big(c_{j+1}^{\dagger}c_{j-1}^{} - \mathrm{h.c.}  \big) -V\big( c_{j}^{\dagger}c^{}_{j-1} - \mathrm{h.c.} \big)(2\hat{n}_{j+1} - 1 ) - V\big(c_{j+1}^{\dagger}c^{}_{j} - \mathrm{h.c.} \big)(2\hat{n}_{j-1} - 1)\Big)			
 \nonumber  \\
  \mathcal{H} = &-& \sum_{j=1}^L  (c_{j}^{\dagger}c_{j+1}^{} +  \mathrm{h.c.} ) + V \sum_{j=1}^L (\hat{n}_{j} - \frac{1}{2})(\hat{n}_{j+1} - \frac{1}{2}) 
	      \frac{1}{2}\sum_{j=l}^{N_{II}} \frac{\sinh \nu_j}{\cosh^2 \nu_j + V^2/4}  \mathcal{H}_j^{\Delta}
  \label{eq:Ham_many_II} 
\eea
\end{widetext}

In order to access larger effective system sizes for bulk systems, Veki\'c and White~\cite{Vekic_1993_1996} introduced the Smooth Boundary Conditions (SBC) that smoothly reduce the energy parameters of the Hamiltonian near the edges of the system. 
Modified boundary conditions were later applied to transport problems with the Damped Boundary Conditions (DBC)~\cite{Bohr_2006}, where the hopping elements where reduced at the
boundaries in a NRG-like fashion. 
The idea of SBC and DBC is motivated by the logarithmic discretization of the Wilson NRG. However, here one uses the exponentially decaying hoppings only as a boundary condition in order to mimic a larger system size.
The DBC are characterized by two parameters: the damping parameter $\lambda$, $t_n = t \cdot \lambda^n$,  and the length or the region where the hopping is reduced, $L_D$.
More precisely, within the $L_D/2$ sites on the right side, the hopping is given by $t \cdot \lambda^{n}$, $n=1 \dots \frac{L_D}{2}$,
and similarly on the right side. 
A restriction of the method is the tuning of the parameter $\lambda$. Indeed, a too strong damping amounts to adding another impurity in the system (the DBC region itself) which induces backscattering and can give questionable values for the conductance.
In this work, we propose a new type of boundary conditions based on integrable impurities that induces no backscattering. 
In the first part, we show that these impurities share the property of enhancing the density of states at the Fermi surface.
The system thus appears to be effectively larger as long as one studies only properties which are dominated by processes at the Fermi surface.
Next we show that the impurities are transparent for wave packets and that this property is preserved in the extension to an interacting chain.

Finally we apply our boundary condition to study the linear transport of an interacting two site impurity via the embedding method.~\cite{Favand_1998, Bohr_2006}
\section{Integrable impurities}
The integrable impurities we are using were previously derived by Schmitteckert \et~\cite{Schmitteckert_PhD,Schmitteckert_1995} using the quantum inverse scattering method~\cite{QISM_book} and the algebraic Bethe ansatz. 
This method makes it possible to determine the eigenvalues of Hamiltonians that correspond to specific scattering matrices which fulfil the Yang-Baxter equation. 
These integrable impurities were derived in the context of the spin$-1/2$ XXZ model which can be transformed to a model of spinless fermions via a Jordan-Wigner transformation~\cite{Jordan_Wigner_1928}. 
Here we are using the simplest type of impurities obtained by this method and one impurity consists of three sites $\{j-1,j,j+1\}$ coupled with the fine-tuned parameters given in the integrable single impurity Hamiltonian $\mathcal{H}_j^{\Delta}$ eq.~(\ref{eq:Ham_one_II}) above.
$V$ stands for the nearest-neighbour 
repulsion in the spinless fermions chain in which the impurities are implemented and $\hat{n}_j$ is the density operator on site $j$. The parameter $\nu_j$ characterizes the strength of the impurity centred on site $j$. 
$N_{II}$ integrable impurities are implemented in an interacting chain of length $L$ with hopping parameter $t=1$ as in eq.~(\ref{eq:Ham_many_II}) above.

In non-interacting leads ($V=0$) such as the ones used in the embedding method, it simplifies as~\cite{Schmitteckert_1995}
\bea
 \mathcal{H}  &=& -\sum_{j=1}^L (c_{j}^{\dagger}c_{j+1}^{} +  \mathrm{h.c.} ) \nn
              &+& \sum_j^{N_{II}} t_j' (c^{\dagger}_{j-1}c^{}_j + \mathrm{h.c.})  
               +  \sum_j^{N_{II}} t_j'' ( e^{i\pi/2} c^{\dagger}_{j-1}c^{}_{j+1} + \mathrm{h.c.}) \nn 
	 t_j' &=& (1 - \frac{1}{\cosh \nu_j}\big) \, \, \, ,  \, \, \, t_j'' =  - \tanh \nu_j
\label{eq:Impurity_Ham_non_interacting}
\eea
Four of these integrable impurities are represented in the boundary region (green) in fig.~\ref{fig:Sketch_systems}.
Setting $\nu_j=0$ removes the impurities and eq.~(\ref{eq:Impurity_Ham_non_interacting}) boils down to a one-dimensional tight-binding Hamiltonian. On the opposite, $\nu_j \rightarrow \infty$ disconnects the site $j$ from the chain and adds a phase $e^{i\pi/2}$ between the sites $j-1$ and $j+1$. 
These impurities break translational invariance and replace it by another symmetry generated by $\nu_j$-dependant operators that can be calculated using higher derivative of the transfer matrices. As a consequence, no degeneracies are lifted, contrary to generic impurities. 
Another noteworthy feature of these integrable impurities is that they induce no localization, as long as the phase on the $(j-1,j+1)$ bonds is $\pi/2$. For this precise value, the squared modulus of the wavefunction of a system in presence of $N_{II}$ impurities 
shows an extended eigenstate with reduced amplitude on the central sites of the impurities. This behaviour is directly related to the transparency of the impurity~\cite{Schmitteckert_PhD,Eckle_1997},
which is in accordance with the findings of \cite{Nilsson_2007}.

\section{Density of states}

\begin{figure}
 \onefigure[width=0.49\textwidth]{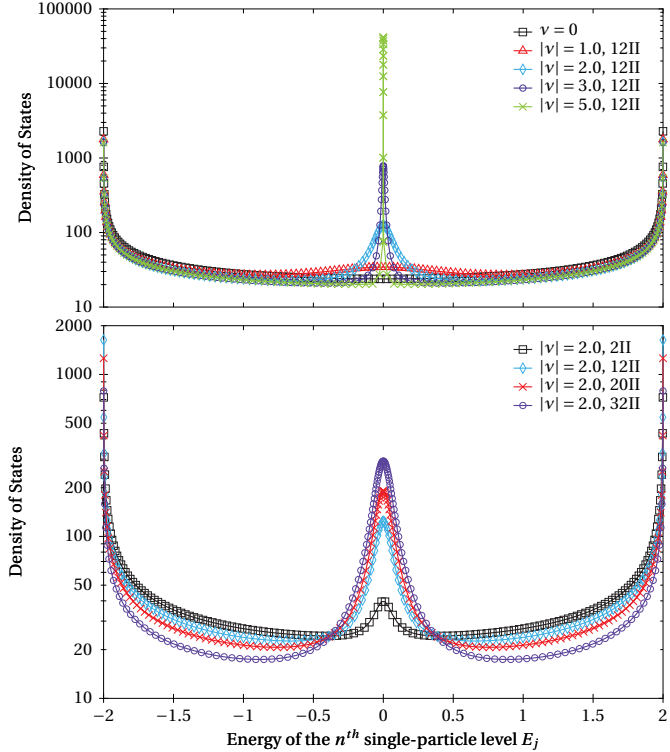}
\caption{(Colour on-line) Density of states obtained by exact diagonalizations in a $L=150$ non-interacting ring pierced by a flux $\Phi=\pi/2$ with $N_{II}$ integrable impurities as boundary conditions ($\nu=\pm |\nu|$). Upper panel: influence of $|\nu|$ at $N_{II}=12$ fixed. Lower panel: influence of $N_{II}$ at $|\nu|=2.0$ fixed.}
\label{fig:Density_of_states} 
\end{figure}
We start by looking at the change of the density of states (DOS), {\it i.e.}\ the inverse level spacing, which is generated in the presence of the integrable impurities.
We performed exact diagonalizations of a non-interacting lead of length $L=150$ pierced by an Aharonov-Bohm flux $\Phi=\pi/2$ to remove degeneracies
of the non-current carrying system.
In the following, we always implement an even number $N_{II}$ of impurities as boundary conditions, 
half of them with a strength of $+\nu$ and the other half $-\nu$ so that the ground state energy is an even function of the phase $\phi$ induced by the Aharonov-Bohm flux. 
The upper panel of fig.~\ref{fig:Density_of_states} shows the influence of the parameter $\nu$ in a ring containing 12 integrable impurities. 
The density of state at the Fermi level increases with $\nu$, leading to an extremely sharp peak at $\nu=5t$.
In the lower panel, $|\nu|$ is fixed at the intermediate value $|\nu|=2$ and we study the influence of the number of impurities $N_{II}$ ({\it i.e.}\ the size if the boundary region). 
The density of states at the Fermi level also increases with $N_{II}$.

\section{Wave packet dynamics}
\begin{figure}
 \onefigure[width=0.49\textwidth]{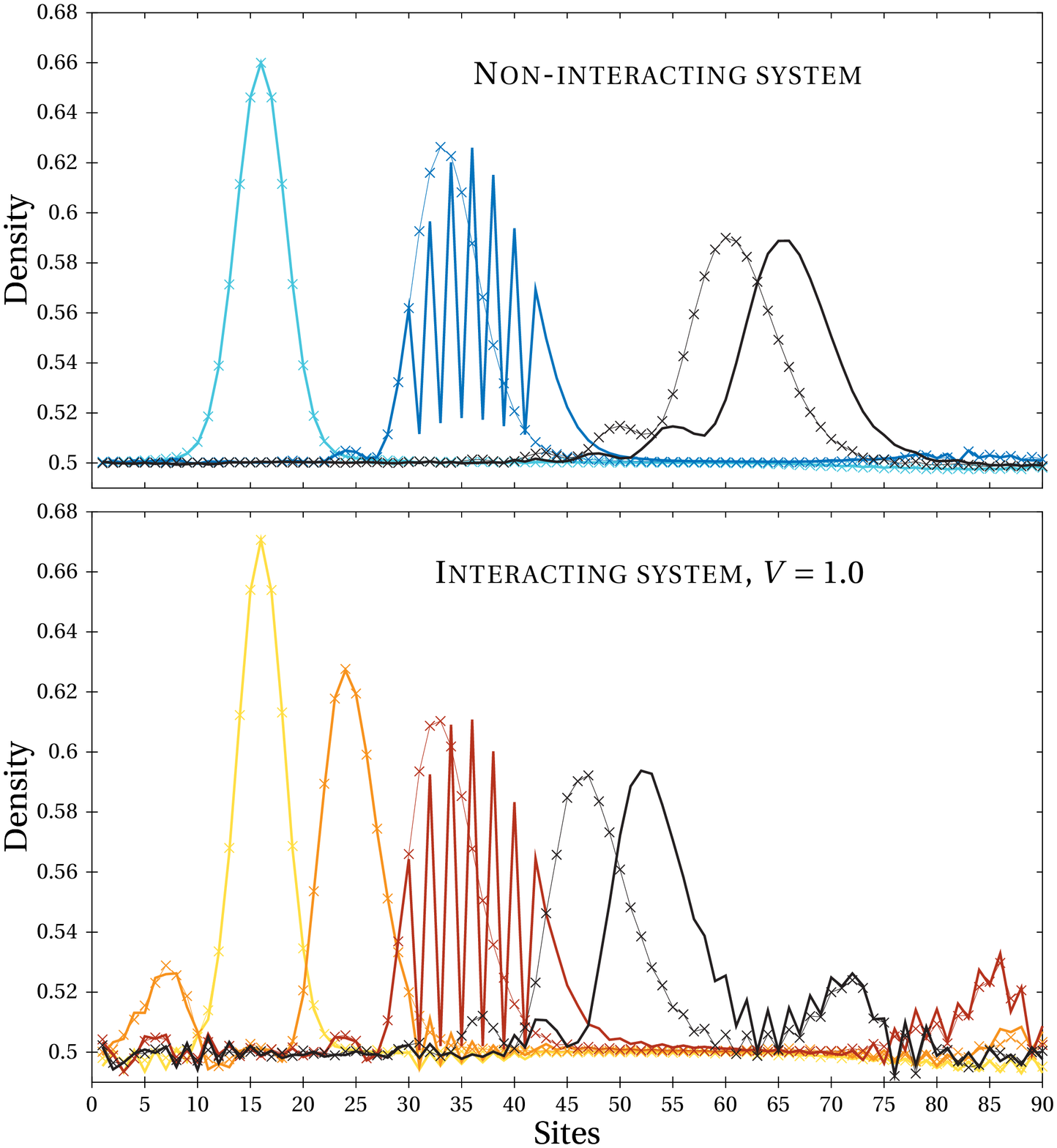}
\caption{(Colour on-line) Snapshots of the evolution of a Gaussian wave packet, starting from site 15, through six integrable impurities ($\nu=1$) centred on sites $j=31\dots41 $ in a non-interacting (top panel) and in an interacting system (bottom panel). For comparison, the curves with crosses and thin lines show the same system at the same time without impurities.}
\label{fig:tDMRG} 
\end{figure}

We now consider the real time dynamics of a wave packet sent on a series of six integrable impurities first in a non-interacting chain of spinless fermions eq.~(\ref{eq:Impurity_Ham_non_interacting}) and then in an interacting Luttinger lead eq.~(\ref{eq:Ham_many_II}) with repulsion $V=1.0$. 
The bulk system is half-filled ($n=1/2$) and the excitation is created by a Gaussian wave packet of right-moving fermions with wave-vector $k=0.6$ ($k>k_F$). 
We use a time-dependant Density Matrix Renormalization Group (td-DMRG) algorithm based on applying the full time evolution operator to an initial state~\cite{Schmitteckert_2004}, 
making 11 sweeps and keeping up to 1000 states in a system with $L=90$ sites and periodic boundary conditions (PBC). 
Fig.~\ref{fig:tDMRG} shows the evolution of the density of the wave packet sent on the series of integrable impurities with $\nu=1.0$ (thick solid lines) and in a system without impurities (curves with crosses) at the same time values. 
Due to band curvature effects the wave packet already acquires a side peak in the homogeneous system. This feature is preserved in the system including the impurities which are indeed completely transparent as expected from the Bethe ansatz integrability of the model.
In case of an interacting wire the wave packet splits into a dominant right-moving part and a smaller left-moving part.
The splitting of the bare electron into left and right-moving low-lying excitations of the Luttinger liquid is due to the density density interaction and the fact, that the fermions themselves are not the bare  constituents of the low-lying excitations. Note that this splitting is also observed  
in simulations of the spin charge separation~\cite{Ulbricht_2009} in one-dimensional Hubbard chains.
Even in the interacting case no additional feature appears in the presence of integrable impurities, which again illustrates the fact that they induce no backscattering. 
However, since the hopping parameters are modified in the impurities, the wave packet acquires a phase that depends on the value of $\nu$ and on the number of impurities it crosses. 
In the limit $\nu \rightarrow \infty$, the wave-packet would travel trough a smaller system of size $L_{\infty}=L-N_{II}$. 
While this result is to be expected from the Bethe ansatz integrability of the model it displays an example where the integrability of the system obtained within an equilibrium description persists in the non-equilibrium dynamics.
Note that in contrast to the NRG tsunami \cite{Schmitteckert_2010} the density of the wave packet is not enhanced in IIBC region, see Fig.~\ref{fig:tDMRG}. 
Thus the IIBC region only leads to an increase of the density of states, while the modification on the velocity is small.

\section{The embedding method revisited}
\begin{figure}
 \onefigure[width=0.4\textwidth]{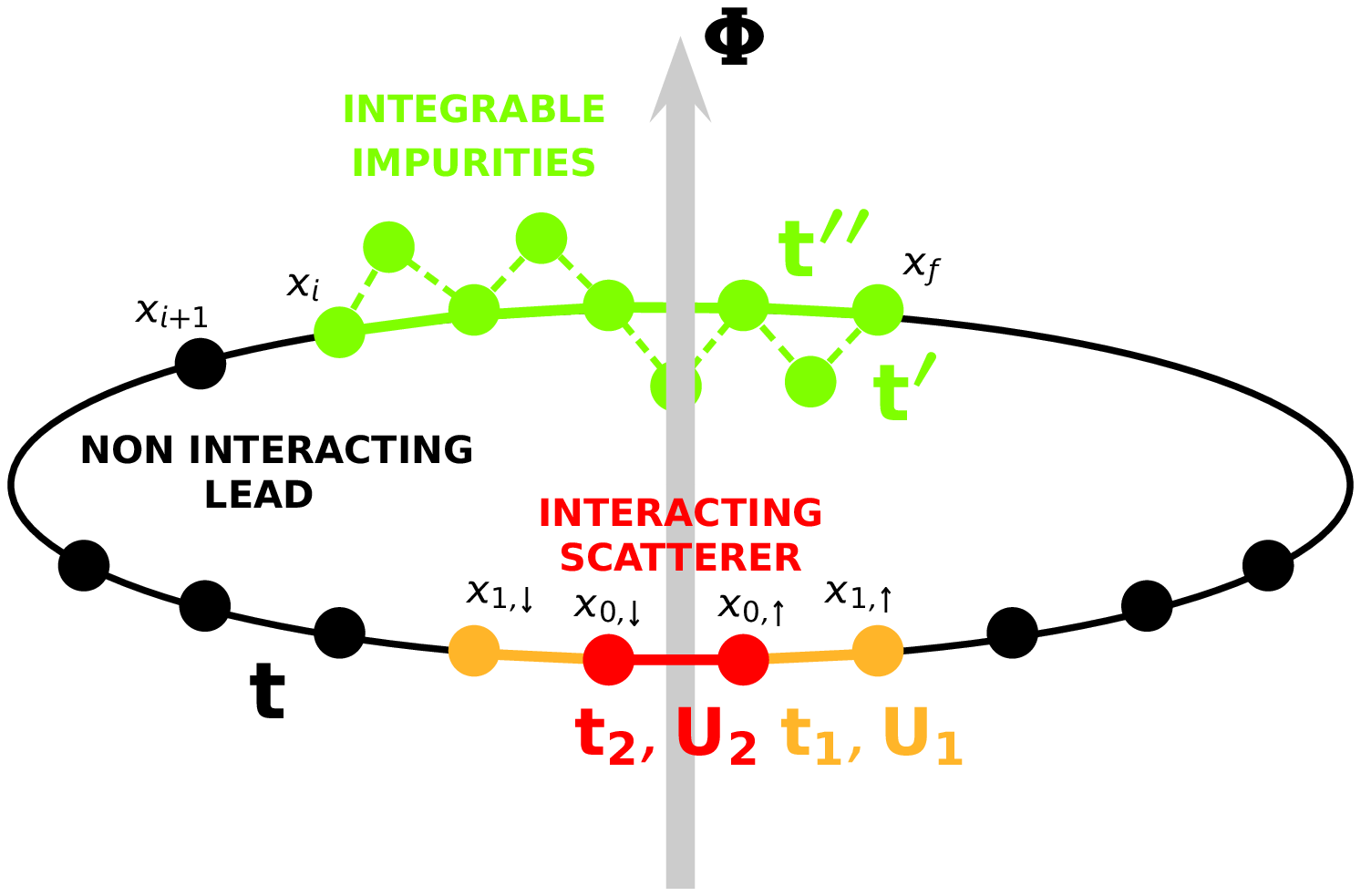}
\caption{(Colour on-line) A 4-sites interacting scatterer embedded in a non-interacting ring pierced by an Aharonov-Bohm flux $\Phi$. The boundary conditions (green) consist in integrable impurities (here four) with parameter $\pm\nu$ (see eq.~(\ref{eq:Impurity_Ham_non_interacting})).}
\label{fig:Sketch_systems} 
\end{figure}

Determining the conductance of low-dimensional strongly correlated structures, such as nanotubes or molecules, is now an experimental and theoretical challenge~\cite{Book_Molecular_Electronics}.
Since it is currently not possible to study a realistic model for transport through molecules including a complete quantum chemistry description using a many-particle approach we restrict ourselves to the study of model systems of interacting structures as benchmarks for more complex systems.
In this section we compute the conductance of a small scatterer, namely a few interacting sites, embedded in a non-interacting tight-binding ring of length $L$ pierced by an Aharonov-Bohm flux $\Phi$ (see fig.~\ref{fig:Sketch_systems}). 
The embedding method~\cite{Favand_1998,Embedding2} is based on the fact that the conductance of the scatterer can be related to the persistent current generated by the Aharonov-Bohm effect, $J=\partial E_0/\partial \Phi$, where $E_0$ is the ground-state energy. 
In a ring of $N$ spinless fermions ($N$ odd \footnote{$N$ even adds one additional term to the current but does not change the physical content presented hereafter.}), with length $L$ and with disorder induced by a general potential $V$, the leading $1/N$ term in the persistent current is~\cite{Gogolin_1994}:
\be
	J(\phi) = -\frac{e v_F}{\pi L}\frac{\arccos\big(|T_{k_F}| \cos \phi \big)}{\sqrt{1-|T_{k_F}|^2\cos^2 \phi}}  |T_{k_F}|\sin\phi 
	\label{eq:current}
\ee
where $\phi=2\pi\Phi/\Phi_0=e\Phi/c\hbar$ and $|T_{k_F}|^2$ is the transmission probability at the Fermi level. 
So far, no analytical solution was proposed for the case of an interacting scatterer. However, in the limit of large electron reservoirs, {\it i.e.}\ leads at least larger than the size of correlations induced in the lead by the scatterer (e.g. Kondo cloud), and under the assumption of Fermi liquid behaviour, the interacting scatterer behaves effectively as a {\it non-interacting} scatterer with an interaction-dependent transmission coefficient~\cite{Sushkov_2001,Molina_2003,Molina_2004}. 
At very low temperature, the conductance of a small interacting region is then obtained by a Landauer-like formula~\cite{Landauer_1957,Buttiker_1986,Meir_1992} $g = |T_{k_F}|^2$.

\section{Study of a Kondo-like model}
We now use these integrable impurities as boundary conditions for the embedding method to study a Kondo-like model and we compare the value of the conductance with the one obtained by other types of boundary conditions. 
We consider a non-interacting tight-binding ring (bulk region) embedding a four-sites interacting region with 
repulsions $U_1$ and $U_2$ and hopping constants $t_1$ and $t_2$, as showed in fig.~\ref{fig:Sketch_systems}. The ring is pierced by an Aharonov-Bohm flux $\Phi$ which is included in the first bond and the modified boundary conditions are applied between the sites $x_i$ and $x_f$:

\bea
 \mathcal{H} &=&   -t \big( \sum_{j=x_i+1}^{x_{1,\down}} + \sum_{j=x_{1,\up}+1}^{x_f} \big) (c_{j}^{\dagger}c_{j-1}^{} +  \mathrm{h.c.} )
	     \nn	
	     &+&  \!\!\!\!\!\! \sum_{j=\{x_{0,\down}, x_{1,\up}\}}\!\!\!\!\!  
                   \Big[ 
                         U_1(\hat{n}_{j-1} - \frac{1}{2})(\hat{n}_{j} - \frac{1}{2}) 
                       - t_1 (c_{j-1}^{\dagger}c_{j}^{} + \mathrm{h.c.} ) 
                   \Big]
	     \nn
              &+& U_2 (\hat{n}_{x_{0,\up}} - \frac{1}{2})(\hat{n}_{x_{0,\down}} - \frac{1}{2}) 
	        - t_2 (c_{x_{0,\up}}^{\dagger}c_{x_{0,\down}}^{} + \mathrm{h.c.} ) 		
 \label{eq:Ham_scatterer}
\eea

By looking at the system before closing left and right leads to a ring, {\it i.e.}\ in a two-lead set-up, we can label the left lead and the first two impurity sites as down spin fermions and the other fermions as up spin fermions. 
By merging the up and down sites with the same distance from the centre into spin 1/2 sites,
$\tilde{c}_{x,\uparrow} := c_x$ and $\tilde{c}_{x,\downarrow} := c_{-x}$,

we can map the model eq.~(\ref{eq:Ham_scatterer}) to an extended single impurity Anderson model, see fig.~\ref{fig:Sketch_systems}.
In this description $U_2$ takes the role of the on-site repulsion, $t_1$ corresponds to the hybridization to the leads, while $t_2$ plays the role of a transverse magnetic field $t_2 (c_{x_0, \up}^{\dagger}c_{x_0,\down}^{} + c_{x_0, \down}^{\dagger}c_{x_0,\up}^{} ) \sim t_2 ( \hat{S}_{0}^{+} + \hat{S}_{0}^{-})$.  $t_2$ has thus a double role in the transport properties. Setting $t_2$ to zero disables any transport, while, as detailed later, a finite $t_2$ reduces and for $t_2 > \TK$ even destroys the Kondo peak. Here we are looking at a regime, where $t_2$ is finite to enable transport, yet small enough to not destroy the Kondo effect. 
The repulsion $U_1(\hat{n}_{1,\sigma} - 1/2)(\hat{n}_{0,\sigma} - 1/2)$ plays a similar role as in the Interacting Resonant Level Model (IRLM) where it broadens the conductance resonance peaks.\cite{Bohr_2007,Boulat_2008}
Hereafter we set $U_2=0.6$, $t_1=0.3$ and $t_2=0.05$ while the lead hopping $t$ is used as an energy scale and $U_1$ varies. 
We studied the effect on the conductance of an asymmetric gate voltage, which in the spin language corresponds to a small magnetic field $B$, to get an approximate value of the Kondo temperature which was found to be $T_K \approx 0.1$.

In the following we use the  DMRG, making 11 sweeps and keeping up to 1700 states, to compute the expectation value of the current 
operator $\hat{J}_j = - t_j \frac{2e}{\hbar} \mathrm{Im} \langle c_j^{\dagger}c^{}_{j-1} \rangle$,

where $t_j$ is the hopping matrix element between sites $j$ and $j-1$.
This value is then fitted with eq.~(\ref{eq:current}) to extract the conductance $g$ and the effective system size $\Leff=v_F/\tilde{L}_{\rm eff}$,
where $v_F$ is the Fermi velocity. 
Note that $\Leff$ should be understood as an effective size only and the results may have deviations from currents measured in a system of
$\Leff$ systems sites in the absence of modified boundary terms due to effects of the Kondo screening cloud~\cite{Affleck_2001}.  

\begin{figure}
 \onefigure[width=0.49\textwidth]{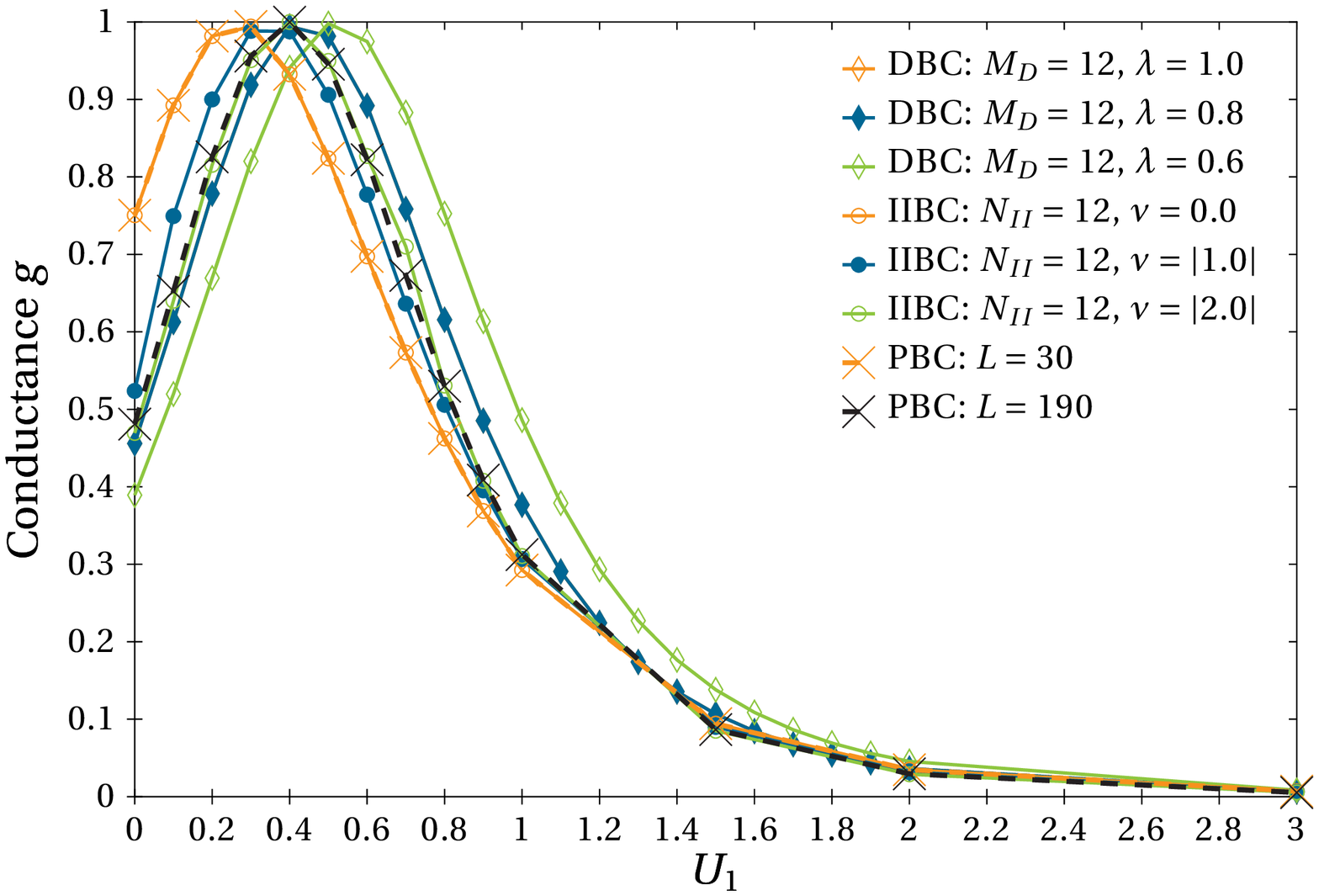}
\caption{(Colour on-line) Conductance $g$ versus $U_1$ for different types of boundary conditions. DBC (diamonds) and IIBC (circles) were applied on systems with $L=30$ for different values of $\lambda$ and $\nu$. For comparison, systems with PBC are also shown (crosses). As expected from the definition of the modified boundary conditions, one recovers exactly the values of the PBC ($L=30$) for $\lambda=1.0$ and $\nu=0$ (orange curves).}
\label{fig:T_versus_U1} 
\end{figure}

Fig.~\ref{fig:T_versus_U1} shows the conductance, extracted from DMRG, versus the repulsion $U_1$ for different types of boundary conditions: PBC, DBC and Integrable Impurities as Boundary Conditions (IIBC). 
DBC (resp. IIBC) were applied on 24 bonds with $M_D=12$ on each side (resp. $N_{II}=12$ integrable impurities) in $L=30$ rings with different values of the damping parameter $\lambda$ (resp. $\nu=\pm |\nu|$). For comparison, rings with $L=30$ and $L=190$ sites and PBC are also shown.
On a qualitative level all three types of boundary conditions provide the same result. We first see a finite conductance, which is reduced from the unitarity limit of $g=2e^2/h$ by the scattering processes introduced by $t_2$. Interestingly, similar observations were made for the IRLM.\cite{Bohr_2007} Interaction on the contact link first increases the conductance, even up to the unitarity limit, while for large $U_1$ the conductance decreases again.
Looking at the quantitative results, the different boundary schemes lead to different results.
\begin{figure}
 \onefigure[width=0.49\textwidth]{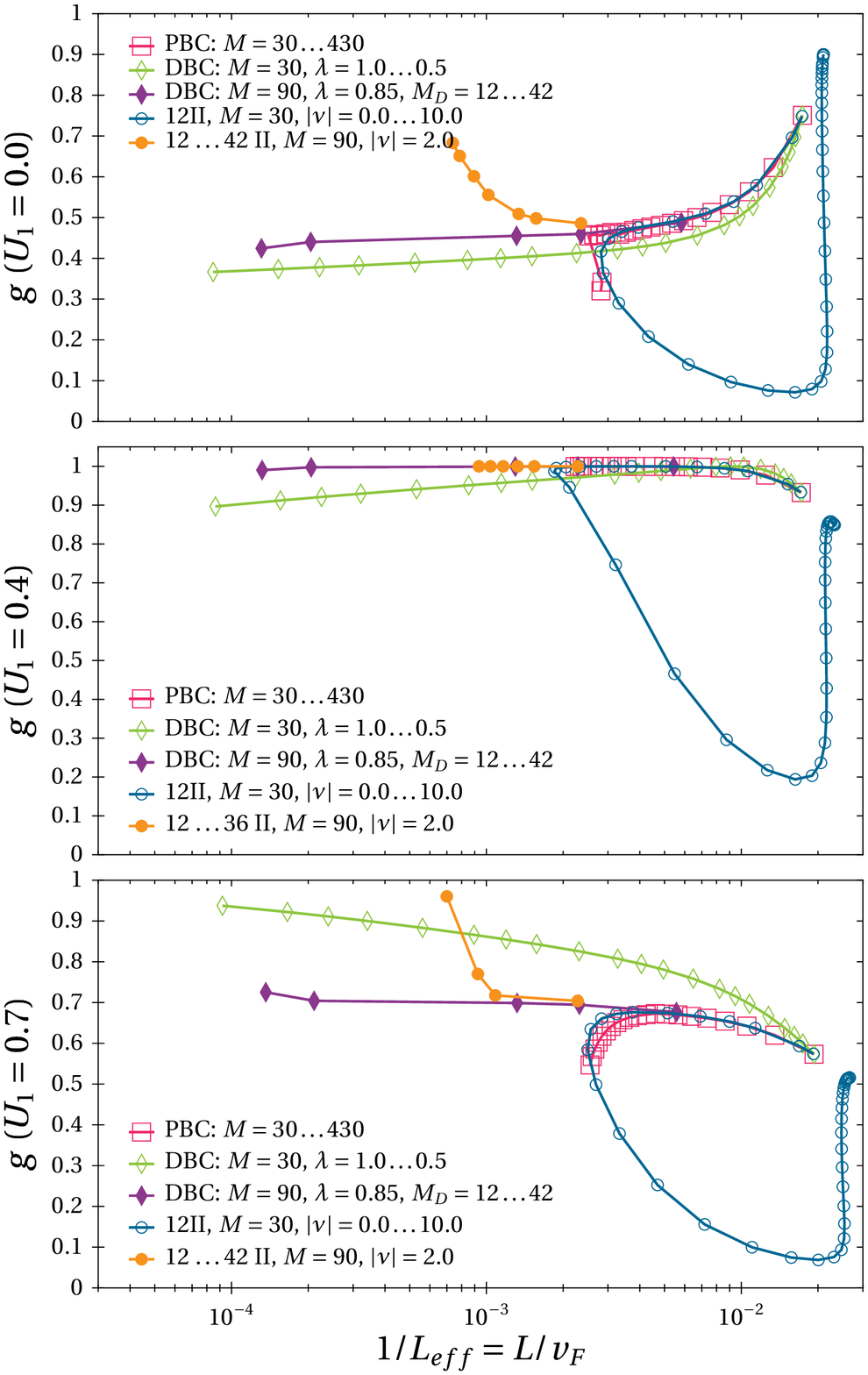}
\caption{(Colour on-line) Conductance $g$ versus $1/L_{\rm eff}$ obtained from eq.~(\ref{eq:current}) for different types of boundary conditions: PBC (squares), DBC (diamonds) and IIBC (circles). Top panel: $U_1=0.0$. Centre panel: $U_1=0.4$. Bottom panel: $U_1=0.7$.}
\label{fig:U01_U04} 
\end{figure}
To elucidate the differences further we compare in fig.~\ref{fig:U01_U04} for three values of $U_1$ ($U_1=0.0, 0.4, 0.7$) 
the conductance versus the effective system size for different types of boundary conditions. 
PBC were applied in systems with up to $L=430$ sites (squares), DBC (diamonds) and IIBC (circles) in smaller systems. 
DBC were applied on 12 sites at each end of a $L=30$ sites rings with varying damping parameters $\lambda=1.0 \dots 0.5$ (green plots). 
On the one hand, it appears clearly in the centre panel of fig.~\ref{fig:U01_U04} ($U_1=0.4$) that for values below $\lambda \approx 0.85$,
 the damping tends to reduce the conductance~\cite{Freyn_2010}, as an effect of backscattering on the boundary conditions. 
For $U_1=0.0$ (resp. $U_1=0.7$), decreasing the value of the damping parameter $\lambda$ from 1 to 0.5 tends to decrease (resp. increase) the conductance as the system size increases. 
On the other hand, the size of the damping region at fixed $\lambda=0.85$ does not have much effect on the conductance (curves with purple filled diamonds)~\cite{Bohr_2006}. 

IIBC were applied on 24 sites ($N_{II}=12$) in $L=30$ sites rings with increasing parameter $|\nu|=0\dots10$. 
A sweeping procedure, where we gradually switched on the $\nu$ parameter, was used in the same way as in~\cite{Bohr_2006} to avoid trapping of the DMRG in some excited states due to the low coupling of the central sites of the impurities for large values of $|\nu|$. 
Fig.~\ref{fig:U01_U04} shows that increasing $|\nu|$ first tends to increase the effective system size, which is the expected effect, and gives values of the conductance comparable with the ones obtained with PBC and DBC. However for $\nu > 2.5$, the conductance and the effective system size strongly decreases, as the central sites of the impurities start to disconnect from the chain. 
The effective system size finally converges to a value related to the size of the ring without the impurities central sites $L-N_{II}$, 
and the conductance saturates.
We also studied the effect of the number of integrable impurities at fixed $|\nu|=2$ (orange plots). 
It appears that increasing the size of the boundary region tends to increase the conductance, eventually leading to a sudden upturn. 
Actually, even for a single weak link without any interaction we find that in the large $\nu$ limit the transmission obtained from the embedding method shows an upturn toward unitary transmission.
We would like to note that the embedding method is restricted to strict one-dimensional leads in order to avoid the problem of channel mixing, for details see~\cite{Freyn_2010}. A system with no scatterer is integrable via the algebraic Bethe ansatz and we expect strict one-dimensional behaviour as given by the analytic solution of the ground state curvature \cite{Schmitteckert_PhD,Schmitteckert_1995}.
However, the scatterer under investigation breaks the integrability of the system which could finally lead to a channel mixing at the impurity and a breakdown of the embedding method. We would like to point out, that this is is not a problem of our integrable boundary condition; at least not as long as one does not extract quantities which rely on a strictly one-dimensional character of the system.

It is important to keep in mind that the conductance should be extracted by performing the thermodynamic limit to infinite lead length.
Ideally, it should be independent from the choice of boundary conditions, as in fig.~\ref{fig:U01_U04} top and middle panel. 
However, as seen in  fig.~\ref{fig:U01_U04} lower panel the conductance obtained from different boundaries conditions can substantially differ, signalling that the result gives only the order magnitude, but not the precise quantitative value. 
By applying different types of boundary conditions, one has a handle to check the sensitivity of a system to these issues. 
We recommend that one should use both types of boundary conditions to check results for consistency instead of relying on DBC and IIBC alone.
\begin{figure}
 \onefigure[width=0.49\textwidth]{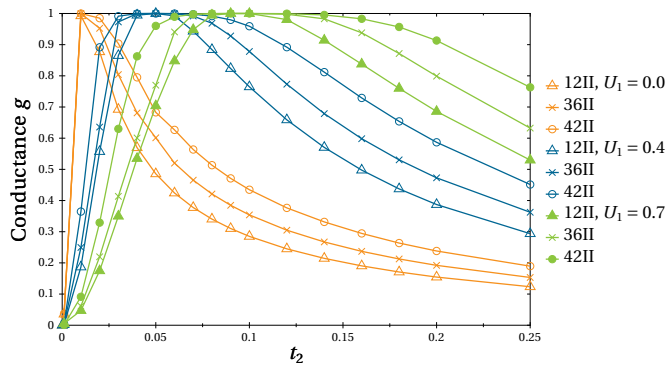}
\caption{(Colour on-line) Conductance $g$ versus $t_2$ for a $L=90$ system with $N_{II}=12$ (triangles), 36 (crosses) and 42 (circles) integrable impurities with $\nu=|2.0|$. The repulsion $U_1$ broadens the region of conductance $g \sim 1$: $U_1=0.0$ (orange plots), $U_1=0.4$ (blue plots) and $U_1=0.7$ (green plots).}
\label{fig:g_vs_t2} 
\end{figure}
Finally, we study the effect of the magnetic field $t_2$ on the conductance at the same three values of $U_1$ (fig.~\ref{fig:g_vs_t2}). 
As expected by the Kondo physics previously discussed, the Kondo peak is destroyed by a magnetic field $t_2 \sim T_K$. Moreover the range of values of $t_2$ for which the conductance reaches $g=1$ 
becomes wider as $U_1$ increases (sharp peak for $U_1=0.0$, wide plateau for $U_1=0.7$) since this interaction tends to broaden the resonance peak. 
The decrease of the conductance for small $t_2$ is related to the shrinking of the resonance width while $t_2$ goes to zero. E.g.\ for $t_2=0$ it has
to be strictly zero. Thus, the unitary conductance values get reduced by our limited resolution stemming from the finite level spacing of the leads.
\section{Conclusion}
We have proposed a new type of boundary conditions based on integrable impurities to increase the effective system size in the frame of the embedding method. 
Unlike other modified boundary conditions, such as DBC, these integrable impurities generate no interfering backscattering. 
We have first justified their use for linear transport problems by showing that they indeed increase the effective system size. 
We have then showed that they are transparent for a wave-packet even in an interacting system. 
We have applied these new boundary conditions to study the transport properties of a system that displays geometric Kondo physics and showed that the Kondo resonance gets broadened by an interaction on the contact link.
We would like to emphasize that the use of the IIBC is not restricted to DMRG. 
It is a very general concept and can be easily implemented in other methods like density functional theories or quantum Monte Carlo techniques.
By making $\nu$ site dependent they can also be implemented within standard or time dependent NRG.

\bibliographystyle{eplbib}

\end{document}